\documentstyle[twoside,fleqn,espcrc2,epsf]{article}

\newcommand{\AmS}{{\protect\the\textfont2
  A\kern-.1667em\lower.5ex\hbox{M}\kern-.125emS}}

\title{
\vspace{-5.0cm}
\begin{flushright}
{\normalsize RU-97-73}\\
\vspace{-0.3cm}
{\normalsize KUNS-1460 HE(TH) 97/13}\\
\end{flushright}
\vspace*{2.5cm}
The overlap passes a chiral dynamical test in two dimensions.
\thanks{
Talk at Lattice '97, Edinburgh, July 22 - 26, 1997.}
}

\author{Yoshio Kikukawa\address{Rutgers University, 
Department of Physics and Astronomy, Piscataway 
NJ08855.}\thanks{Permanent address: 
Department of Physics, Kyoto University,
Kyoto 606-01, Japan.} and 
Herbert Neuberger${\hskip .07em\relax}^{\rm a}$\thanks{Presenter; 
Research supported in part by the
DOE under grant \#DE-FG05-96ER40559.}}

\begin{document}

\begin{abstract}
A certain $U(1)$ model in 2 dimensions, 
describing four right handed unit charged
Weyl fermions interacting with one 
doubly charged left handed Weyl fermion,
is exactly soluble and has massless 
Majorana--Weyl composites. Instanton
induced fermion number violation is 
essential for 't Hooft anomaly
consistency. The associated 't Hooft vertex can be 
analytically computed in the continuum, 
including finite size corrections. 
This number is reproduced in a dynamical 
simulation employing the overlap.
\end{abstract}
\maketitle
This contribution is based on joint work \cite{NN,YNN1,YNN2}, but is
written in first person, 
close to the oral delivery by the
presenter.

In this talk 
I focus on continuum physics and on the central role
played by gauge field topology in our model. 
I shall introduce a certain
one-point function, a dimensionless 
vertex $V(x)$, whose expectation value
has been analytically computed, finite size effects
included. $\langle V \rangle$ has 
been reproduced in a full dynamical
simulation based on the overlap.

The model in the abstract could be called the 
11112 model, or the 412 model.
It is in two (Euclidean) dimensions, 
has gauge group $U(1)$, and the following
fermionic matter content: 4 RW fermions 
with $q=1$ ($\chi_i , i=1,2,3,4$)
and one LW fermion with $q=2$ ($\psi$). 
R(L)W means right(left)
moving Weyl fermions. The model can be 
generalized to the $SU(2)$ gauge group
with matter content of 2 RW of $I={1\over 2}$ 
and 1 LMW of $I=1$.
MW means Majorana-Weyl.

The model is interesting because:
\begin{itemize}
\item It is definitely chiral (I mean it is 
not vector like in disguise - this is evident as 
it has an odd total number of W fermions).
\item There is confinement, a trivial 
consequence of two dimensional
electrodynamics. 
\item The gauge field configurations fall
into separate topological classes and there
are instanton solutions on a toroidal
space-time. 
\item The model is gauge invariant by {\it nontrivial} anomaly
cancelation; $2^2 = 1^2 +1^2 +1^2 +1^2 $.
\item The massless spectrum consists of 
composite ``baryons''. These ``baryons''
are neutral, made out of two 
distinct $\chi_i$'s and one anti-$\psi$. 
It follows that they are massless 
right movers. Naively, one
would guess that the baryons are 
W fermions, but this is wrong:
The anti-baryon can mix with the
baryon despite the difference in fermion numbers and the baryons
are MW. 
(The particular 11112 charge 
values ensure a particle interpretation
of the massless sector of the 
theory. Generically,
the massless sector of the full theory would be 
described by a
conformal two dimensional theory 
admitting no particle interpretation.)
\item The model contains a nontrivial example of
't Hooft matching. The symmetry in question 
is a global $SU(4)$ acting on the $\chi_i$'s which
transform as a ${\underline 4}$, 
the fundamental representation. Under the
same $SU(4)$ the baryons transform in the six dimensional 
(antisymmetric) representation. 't Hooft matching requires the
baryons to be MW fermions, not W fermions. 
\item The model is exactly soluble.  
't Hooft matching only provides a consistency check 
on a proposed particle spectrum. Here the particle
spectrum can be directly found. 
For the spectrum to come out correctly, in any regularization,
it is imperative that fermion number violation of
the exactly right amount occur in the continuum limit.
\item Exact solubility extends to a finite torus. 
We want to get real numerical agreement for some quantity
measuring fermion number violation - if agreement
is not found the spectrum will not be reproduced. 
It would be quite difficult to
get good numbers 
if we also had to fight numerically against finite size
effects, since the theory has massless particles. It is
therefore almost a necessary requirement in practice, 
at least at the beginning, when suspicions about any proposal
ought to be high, to be able to
provide analytic results that take finite volume effects
fully into account in the continuum. 
\item In the continuum the chiral determinant is positive. 
The road to simulations of truly chiral models
is paved with difficulties. It is enough not to surmount one
and a serious positive test gets out of reach. 
A major obstacle to Monte Carlo
simulation is the expected complex integration
measure. Thankfully, here, in the continuum, on the torus,
there is a set of boundary conditions that make the
product of chiral determinants positive. On the
lattice this is not exactly true, but the
phase is therefore just a lattice artifact,
and one can get away without importance sampling, including
the phase (and most of the real part) in the observable.
The positivity is a special property of our model,
but without it, full dynamical simulation would
be much more difficult. As stressed before, there
is no doubt that the model is genuinely chiral.
\end{itemize}

The dimensionless 
fermion number violating 't Hooft vertex we chose to
work with is given by:
$$
V(x) = {{\pi^2}\over {e_0^2}} \chi_1 (x) 
\chi_2 (x) \chi_3 (x) \chi_4 (x) \bar\psi 
(x) (\sigma\cdot\partial ) \bar\psi (x) .
$$
We derived an exact analytical expression 
for $\langle V\rangle_{t\times l}$  
on a torus of sides $t$ by $l$. 
Our simulation parameters 
need $\langle V\rangle_{t\times l}$ 
at $t=l={{3\sqrt{\pi}} \over {2e_0}}$ ($e_0$ is the continuum
gauge coupling) and there $\langle V\rangle_{t\times l}= 0.0389$, 
{\it substantially} smaller than the value at infinite volume.

The objective of our numerical work is to 
reproduce this last continuum number on
the lattice. By simple arguments 
we expect convergence as $a^2$,  
where $a$ is the lattice spacing, 
defined as the ratio between the physical
size of the lattice and the number 
of sites used, $L^2$. One therefore
expects convergence as $1/L^2$ 
towards the continuum result. 
Within the statistical accuracy of 
our results this indeed happens (see figure). 
\begin{figure}[htb]
\epsfysize=2.7in
\centerline{\epsffile{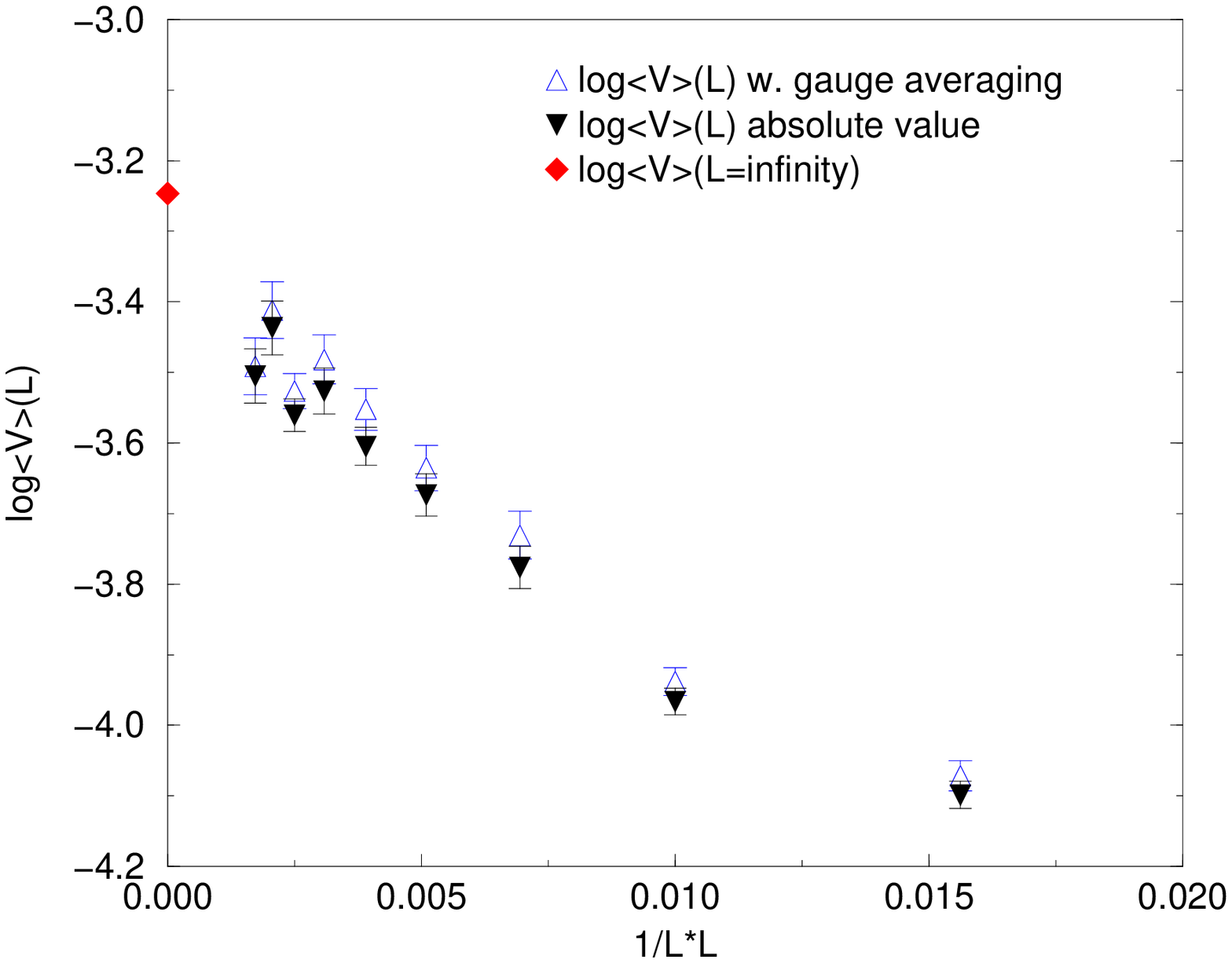}}
\end{figure}

We worked with  one point functions,
since these typically have the least statistical noise and
it was important for us to get good numerical agreement.
There is one possible source of
contamination that we cannot eliminate cleanly,
namely the generation of a marginal Thirring coupling. 
Such couplings were seen to be generated in single and 
multi flavored vector like
Schwinger models regularized by the overlap \cite{nnv}, so their
presence here is not a consequence of chirality. 
The evidence that the
numerical effects we observed in the vector like case
(where simulations are much easier)
indeed were Thirring couplings is based
on the dependence on the number of 
flavors and the lattice spacing. 
To avoid the generation
of Thirring couplings we tune a free 
mass parameter in the overlap.
This parameter is an extra ultraviolet cutoff, beyond
the lattice spacing. 

Let me end with some closing comments of a more general
nature. Below are three  very basic questions that 
would appeal to any particle physicist,
working on lattice field theory or not. 
A serious proposal for a general scheme to
non-perturbatively regularize chiral gauge theories 
should answer positively all three of them.

\begin{enumerate}
\item Does the proposal hinge on 
anomaly cancelation ? This excludes
any ``quenched'' tests or mean 
field arguments: one loop fermion
physics is essential.
\item Does the proposal collapse to a 
simple clear lattice formulation
when applied to a vector like theory ? 
Too often is it unclear whether 
new proposals work at all in the vector like case.
\item Does the proposal address 
directly and explicitly instanton induced
violations of otherwise good conservation laws ?
\end{enumerate}
It is inconceivable to me that a 
general scheme could work in four
dimensions but fail in two. Thus, 
I would add to the above minimum 
a fourth question:
Does the proposal work in a
nontrivial two dimensional case, similar to the 412 
model above ? We have done 
a certain amount of ground work in the
continuum which could be useful also
to others who would choose specifically the 
412 model as a test case for
their proposal. 

The overlap should not be the 
single way to regulate chiral
fermions. However, at this juncture in time, it has
established a sort of a benchmark 
for what a proposal can do, and
this progress should not be ignored.

I think there are two major 
levels at which there are open
issues. At the most fundamental level 
lies the question of
existence of an asymptotically 
free chiral gauge theory 
(this would include the 
supersymmetric case). Some (e.g. \cite{mont})
conjecture these theories do not 
exist. Others (e.g. \cite{thooft})
try to rigorously prove they do. 
The fundamental issue is sharp,
a question in mathematical physics, 
independent of what Nature chose
to do. The second level is really 
relevant only if one believes
that there do exist asymptotically 
free chiral gauge theories,
since one cannot prove a negative, 
and the collection of failures
is large enough already. The issue 
is what constitutes, by
present day standards, a credible, 
valid approach. I claim that
a positive answer to the three questions 
is a minimum requirement.
The overlap satisfies this minimum 
requirement and went beyond it.
I invite other proposals to attain
a similar stage. 

Having emphasized more stringent requirements I should also
stress that one should not err in the opposite direction.
While numerical QCD research is of potential
importance phenomenologically, it would be a mistake
to require workers on lattice chiral gauge theories
to produce now results at the lattice QCD level, because this
would simply shut down the subfield and any progress
would stop. QCD work
might \cite{nex}, and already has \cite{nv}, benefited from 
research on lattice chiral gauge theories. 
And we should never forget that
the problem of regularizing chiral gauge theory is of a fundamental
importance exceeding that of determining accurate
numerical predictions of QCD.

\vfill\eject

\end{document}